\documentclass[fleqn,usenatbib]{mnras}

\usepackage{mathptmx}
\usepackage{txfonts}
\usepackage{xcolor}
\usepackage{ulem}

\usepackage[T1]{fontenc}

\DeclareRobustCommand{\VAN}[3]{#2}
\let\VANthebibliography\thebibliography
\def\thebibliography{\DeclareRobustCommand{\VAN}[3]{##3}\VANthebibliography}

\usepackage{graphicx}	

\usepackage{diagbox}

\title{Reanalyzing Large-Scale Structure Using an Updated Gamma-Ray Burst Spatial Density Approach
}

\author[I. Horvath et al.]{
Istvan Horvath$^{1}$,
Zsolt Bagoly$^{1,2}$,
Jon Hakkila$^{3}$,
Lajos G. Balazs$^{4,5}$,
Janos Horvath$^{6}$,
\newauthor 
Sandor Pinter$^{1}$,
Istvan I. Racz$^{1}$,
Peter Veres$^{7,8}$, 
and 
L. Viktor Toth $^{5,9}$\thanks{E-mail: toth.laszlo.viktor@ttk.elte.hu}
\\
$^{1}$Department of Natural Science, University of Public Service, Budapest, Hungary\\
$^{2}$Department of Physics of Complex Systems, E\"otv\"os University, Budapest, Hungary \\
$^{3}$Department of Physics and Astronomy, The University of Alabama in Huntsville, Huntsville, AL 35899, USA  \\
$^{4}$Konkoly Observatory, Research Centre for Astronomy and Earth Sciences, Budapest, Hungary\\
$^{5}$Department of Astronomy, E\"otv\"os Lor\'and University, Budapest, Hungary\\
$^{6}$Visionary Tech \& Event Solutions, Sacramento, California, USA\\
$^{7}$Department of Space Science, University of Alabama in Huntsville, Huntsville, AL 35899, USA\\
$^{8}$Center for Space Plasma and Aeronomic Research, University of Alabama in Huntsville, Huntsville, AL 35899, USA\\
$^{9}$University of Debrecen, Faculty of Science and Technology, Egyetem tér 1, H-4032 Debrecen, Hungary\\
}

\date{Accepted XXX. Received YYc; in original form ZZZ}

\pubyear{2026}

\begin{document}
\label{firstpage}
\pagerange{\pageref{firstpage}-\pageref{lastpage}}
\maketitle

\begin{abstract}

In the past few decades, large universal structures have been found that challenge the homogeneity and isotropy expected in standard cosmological models. 
This study examines burst clustering in both galactic hemispheres using a recently-developed methodology, 
using spheres in 3D space for testing regularities.
Using our new method in both hemisphere we find only one deviation from isotropy.
A small one in the southern and a huge one in the northern hemisphere.
This itself suggests that the two deviations do not likely to come from statistical fluctuation.
The northern huge group contains app. 125 Gamma-ray bursts (GRBs) corresponds with the so-called Hercules-Corona Borealis Great Wall.
The southern group contains 4--5
GRBs locating very close to each other. Two of them (GRB050822 and GRB050318) are  close not just in redshift and the angular position but they are very close in observing time (5 months). 
We concluded that the third important result of this work that this method could not find other  overdensity deviation from homogeneity in the GRBs spatial distribution.
We have shown that the large-scale density increase in the spatial distribution of gamma-ray bursts does not necessarily violate the cosmological principle.

\end{abstract}




\section{Introduction}

Gamma-ray bursts (GRBs) are brilliant transient events observed through their emission of very high energy gamma-ray and x-ray photons \cite{galaxies13010002,2024Univ...10..260F, 2025RAA....25b5001Y}. GRB emission accompanies the relativistic outflow of material expelled during the formation of stellar black holes created in collapsars and in merging neutron stars \cite{galaxies10030066,racz_2018an}. The tremendous luminosities associated with GRBs allow them to be observed at cosmological distances.  When taken in conjunction with their origins in or close to star formation regions, high luminosities allow GRBs to serve as tracers of star-forming matter, and therefore also of large-scale universal structure \cite{Wang15,2016SSRv..202..195P,racz_akarifis_marton_iau_2016}. Large-scale structure represents anisotropies and heterogeneities in the otherwise homogeneous distribution of matter hypothesized by the Cosmological Principle \cite{2000ApJ...536....1L,Shawqi_Al_Dallal_2024}. Large anisotropies have previously been reported in the distributions of galaxies and galaxy clusters, voids, quasars, and gamma-ray bursts ({\em e.g.,} \cite{Gott05, 2020ApJ...897..133P, 2023MNRAS.519L..45S, 1991MNRAS.249..218C, 2012MNRAS.419..556C, clo12, 2022MNRASLopez, 2024JCAP...07..055L, BalazsRing2015,hbht15}).

Large-scale structure is routinely mapped with high-statistics galaxy and quasar surveys \citep{Gott05, clo12}, which provide orders-of-magnitude larger samples and relatively uniform selection across their survey footprints and thus yield stringent constraints on clustering. 
We use GRBs not as a substitute for these approaches, but as an independent and complementary tracer.
Long-duration GRBs arise from the gravitational collapse of massive stars. This physical origin allows them to effectively probe star-forming galaxies. This population differs significantly from quasars. Quasars typically reside within massive host galaxies and high-density environments associated with AGN environments.
GRBs are detected in gamma-rays over wide regions of the sky and can be observed to very high redshift; their afterglow spectra often give spectroscopic redshifts that provide  precise radial information. 
Analyzing large ($\approx$ Gpc scale) structures requires broad angular footprints and significant depth in redshift. The current GRB redshift catalog meets these specific needs. Although the sample is sparse, its wide-area coverage remains a distinct advantage. It serves as atool for examining the large-scale distribution of matter, good for testing cosmic isotropy and homogeneity.  
Our findings provide a tool to test reported GRB overdensities with an improved three-dimensional statistic.

Although GRBs can be seen to very large distances and long lookback times, the immense size of the universe requires a fairly large sample size in which heterogeneities indicative of large-scale structure might be observed \cite{li2015,2012grb..confE..82G}. GRBs have been detected by orbital satellites at a rate of about one per day since the early 1990s. Over the past 35 years, this means that around 12,000 GRBs have been or could have been detected. This is a fairly sizeable potential sample \cite{2024A&A...685A.163L}. However, not all of the bursts that might have been detected are usable for studying large-scale structure. Due to varying instrumental trigger limits, many bursts have been too faint to formally trigger instrumentation or to accurately located detected bursts on the plane of the sky. Most importantly, only a small percentage (app. 5\%) of GRBs have been associated with delayed emission in the form of x-ray/visual/infrared/radio afterglows \cite{2025MNRAS.542..215L}. Afterglow observations are critical because redshifts obtained from them provide a mechanism by which the radial component of GRB locations can be measured.  

The BeppoSAX and Swift satellites ushered in a new era of GRB study by recognizing that bursts were generally followed by lower-energy afterglows \citep{1997Natur.387..783C}. Rapid follow-up of the x-ray afterglow allowed bursts to be localized with significantly better precision than had previously been available, and visual and infrared observations of the fading afterglow also often resulted in redshift measurements and identifications of host galaxies. During the Swift era (which started in 2005 and continues to the present day), GRB detections have been regular, GRB localization uncertainties have been small, and GRB redshift measurements via afterglow observations have been frequent. The percentage of GRBs with known distances has increased, leading to a corresponding increase in the number of GRBs that can be used in cosmological mapping.

Previous studies of large-scale structure using GRBs have identified several statistically-significant GRB clusters. The largest of these is the Hercules–Corona Borealis Great Wall (HerCrbGW), a 2-3 Gpc clustering of northern galactic hemisphere GRBs 
\citep{hhb14, hbht15, HSZ20}. The Giant GRB Ring \citep{BalazsRing2015, BalazsTus2018} spans a redshift range of 0.78<z<0.86 and measures approximately 1.72 Gpc in size. Both of these structures are significantly larger than the theoretical limit of 356 Mpc predicted by the standard cosmological model for the scale of homogeneity \citep{Yad2010} and thus highlight the importance of studying GRBs as probes of large-scale universal structure.

The primary strategy for identifying GRB clusters has been to partition the known distribution into radial (redshift) ranges, then search each radial (redshift) range for two-dimensional concentrations of bursts. Cluster significance is obtained by a variety of statistical clustering tests, each of which assumes independence between the radial (redshift) information and the angular information on the plane of the sky. In this manuscript, we present an alternative approach that searches for clusters by simultaneously using the angular and radial information.

The paper is organized as follows. In Section 2, we describe the GRB dataset and the analysis method used in this study. In Section 3, we present the results for the Northern galactic hemisphere, followed by the findings for the Southern galactic hemisphere in Section 4. We discuss the implications of our results in Section 5 and provide a summary in Section 6.

\section{Methodology}

\subsection{GRB Dataset}

Our dataset consists of 542 GRBs with accurately measured redshifts and angular coordinates 
on the celestial sphere. These GRBs originate from spectroscopic observations, ensuring reliable distance estimates crucial for spatial analysis. They have been detected primarily by NASA's Swift and Fermi missions, with supplementary data acquired from publicly accessible catalogs, including the Gamma-Ray Burst Online Index (GRBOX) and Jochen Greiner's compilation (see Section Data Availability).

\begin{figure}
    \centering
    \includegraphics[width=0.97\columnwidth]{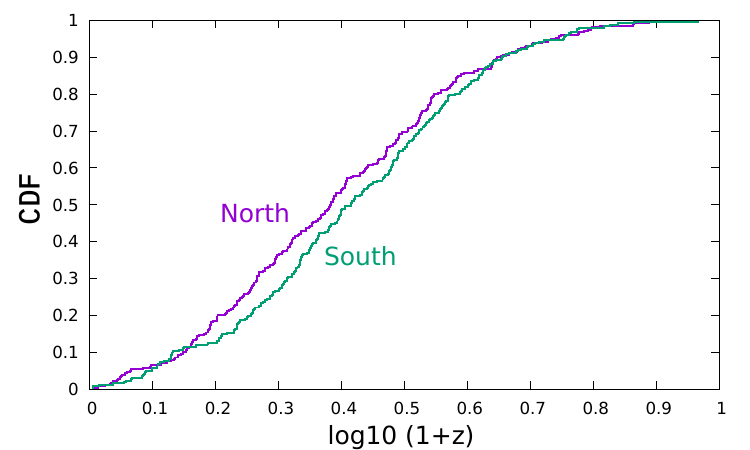}
    \caption{The cumulative distribution function (CDF) of $\log_{10}{(1+z)}$ for GRBs in this study. Purple indicates the 262 GRBs found in the Northern Galactic sky, while green indicates the 280 GRBs found in the Southern Galactic sky. The difference between the two CDFs is not significant.}
    \label{fig:cumdist}
\end{figure}

\begin{figure}
    \centering
    \includegraphics[width=0.97\columnwidth]{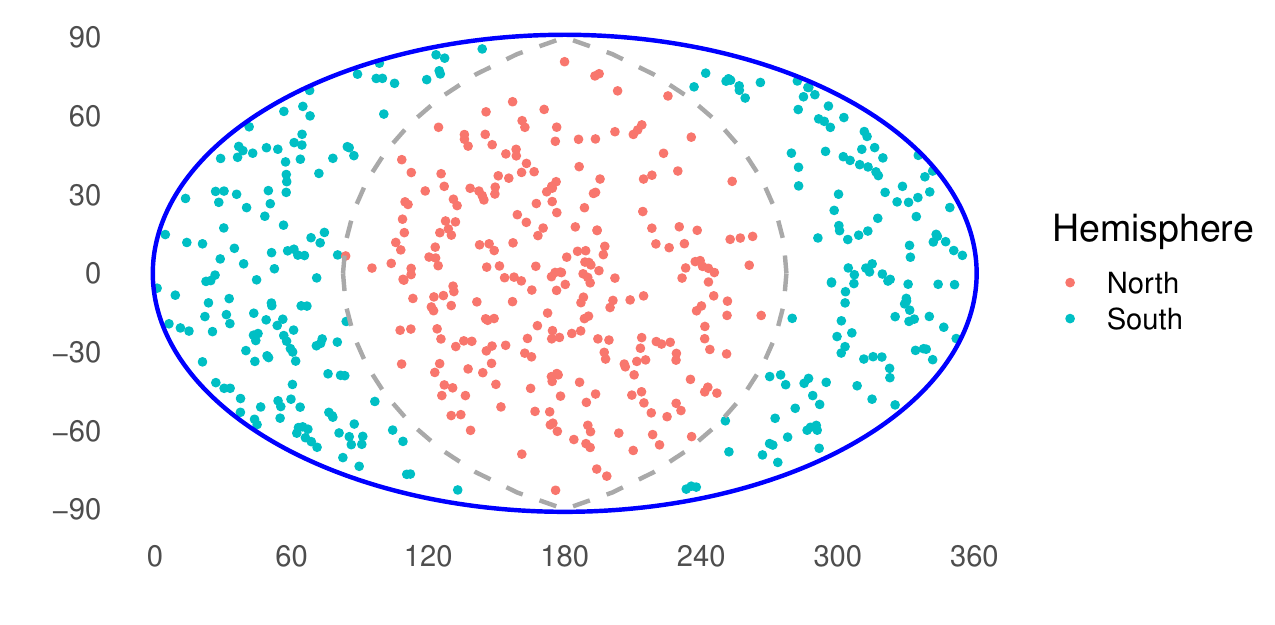}
    \includegraphics[width=0.97\columnwidth]{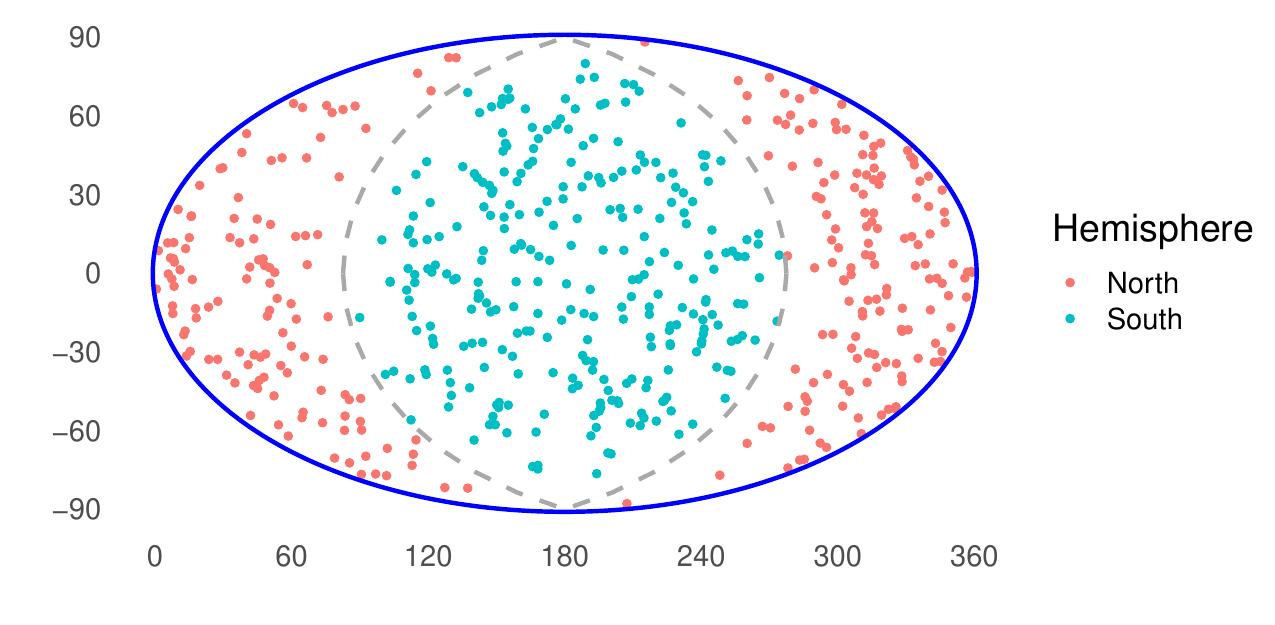}
    \caption{The sky distribution of the 542-GRB sample viewed from the Galactic poles. The top/bottom figure is centered on the Northern/Southern Galactic Pole. The grey dashed line represents the location of the Galactic equator.}
    \label{fig:galacticPoleincenter}
\end{figure}

Given the variability in redshift measurements across different instruments, we apply a quality threshold for redshift inclusion, ensuring that only GRBs with well-documented spectroscopic data contribute to our clustering analysis. This avoids systematic biases due to uncertain photometric redshift estimates and ensures a consistent and robust dataset.
Fig.~\ref{fig:cumdist} and ~\ref{fig:galacticPoleincenter} show the redshift and sky distribution of the 542 GRBs.

Selection effects are the basic limitation factor when using the subset of GRBs with measured redshifts.
The probability of high-energy detection, subsequent ground-based follow-up, and successful spectroscopic redshift assignment is a complex function of celestial coordinates, logistical resource allocation, mission-specific heuristics, and the inherent decline in spectroscopic sensitivity at high redshifts, see \citet{2012A&A548L7Tello} and  figure 1 in  \citet{2025ApJ989.161Lien}.  
The current analysis evaluates spatial clustering against synthetic null distributions generated through volumetric resampling, calibrated to emulate the empirical sky coverage and longitudinal redshift profiles.
We report results separately for the Northern and Southern hemispheres. This strategy mitigates sensitivity to smooth observational gradients, but it does not constitute a full (instrument and follow-up) selection function.
Any significant overdensity should be regarded as a candidate structure pending dedicated exposure/completeness modelling and cross-checks with galaxy/quasar surveys in the same redshift range.

To minimise sensitivity to celestial selection effects --- including reduced GRB/afterglow visibility near the Galactic plane and uneven follow-up due to different observational constraints --- we do not assume an isotropic sky distribution for the null hypothesis. 
In our Monte-Carlo catalogues we keep the observed angular coordinates fixed and randomise only the radial coordinate by resampling/shuffling redshifts within the sample.
This procedure randomly assigns redshifts to GRB positions. Consequently, nonrandom density  enhancements  disappear in (l,b,z) space, after performing MC. 
This methodology, detailed in Sect.~\ref{sec:method}, ensures that our synthetic catalogues will show the same specific observational constraints, providing a tool where large-scale structures can be statistically evaluated.
Therefore the resulting statistical significance is sensitive only to the real  three-dimensional clustering.

A substantial fraction of GRBs with measured redshifts in our compilation were detected by the Swift BAT. It has a non-uniform sky exposure driven by the satellite pointing history and an angle-dependent sensitivity function\citep{2004ApJ...611.1005G,2005SSRv..120..143B,2012A&A548L7Tello}; for the current Swift exposure see Fig.~1. of \citet{2025ApJ989.161Lien}. 
Our analysis take care the observed angular positions during the construction of the MC catalogues (Sect.~\ref{sec:method}). Thus any Swift-like exposure pattern was preserved in the MC realisations making the reported significances be an excess clustering beyond the observational biased density.

The probability of measuring a proper spectroscopic redshift is not uniform on the sky, because the detectability of key emission and absorption features depends on wavelength coverage, spectral resolution and signal-to-noise \citep{2011A&A...526A..30G,Schulze15}. 
Because our randomised catalogues are constructed by shuffling/resampling redshifts from the observed sample (Section~\ref{sec:method}), any global redshift-dependent redshift-success function is propagated into the MC distribution.
However, redshift success may also couple to sky position through heterogeneous follow-up resources; our North/South split partly addresses this, but a full treatment would require explicit completeness modelling which is not yet available.

While the current approach handles large scale selection effects (e.g. galactic extinction), it remains sensitive to smaller scale dependencies where the spatial mask may be intrinsically coupled with the radial distribution. Consequently, any variations in the redshift-dependent selection function that correlate with angular regions might mimic physical clustering. Hence it is necessary to cautiously interpret any signal in the presence of complex incompleteness. Using e.g. full Swift BAT or Fermi GBM catalogues one can check for any high energy parameter for such directional anisotropy, but these searches (e.g. \cite{2017ApJ...851...15R,2019MNRAS.486.3027R,2022Univ....8..342B}) were not able to show any effect. The full selection and optical pipeline is probably more complicated, showing strong human induced effects (e.g. \cite{2014styd.confE..60B}, Fig.~7. and discussion in Sect~4.2. of \cite{2022Univ....8..342B} and Fig.~1. and discussion in Sect.~4. of \cite{2025AcPol..65....9B}), probably these are not fully reproducible.

The present framework effectively marginalizes over the primary selection functions by conditioning the null hypotheses on the empirical angular distribution and precisely replicating the global redshift distribution. However, this methodology does not fully preclude a case where the survey sensitivity might vary simultaneously across both the celestial coordinates and the radial distance. The independence of sky and radial components was investigated with Mann-Whitney U-test by \cite{2022Univ....8..342B}. Although a localized "Faraway GRB Patch" showed a marginal deviation from randomness ($\approx 1$\% significance), using the whole dataset the correlation between redshift and celestial position was excluded. The spatial two-point correlation estimations based on the analysis also remained consistent with zero.

\begin{figure}
    \centering
    \includegraphics[width=0.97\columnwidth]{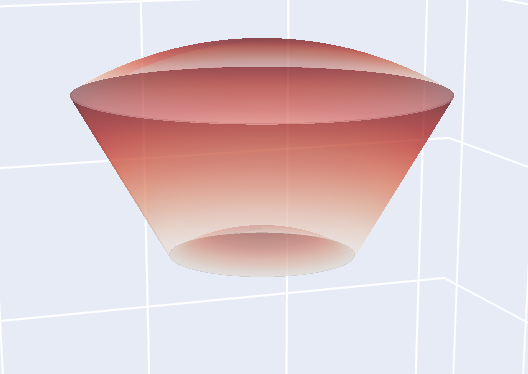}
    \caption{The shape of the previously used volume in the literature. With redshift ($z_1 < z < z_2$) and angular (a circle in the sky) limits.}
    \label{fig:dugo}
\end{figure}

In this study, we aim to provide a more comprehensive analysis of the spatial distribution of gamma-ray bursts by investigating their clustering within well-defined three-dimensional volumes. Unlike previous studies that relied on ad hoc angular and redshift constraints (see Fig.~\ref{fig:dugo}), we now analyze GRB density within spherical volumes centered at specific locations in space (local volumetric resampling).  
This methodological refinement ensures a systematic assessment of cosmic structure using GRBs as tracers of the large-scale distribution of matter in the universe. Our approach builds on previous work of \cite{horv24MNRAS} (Horvath et al. 2024, hereafter referred to as mnras2024), expanding upon the framework used to identify GRB overdensities.

\subsection{Analysis method}
\label{sec:method}

\begin{figure}
    \centering
    \includegraphics[width=0.97\columnwidth]{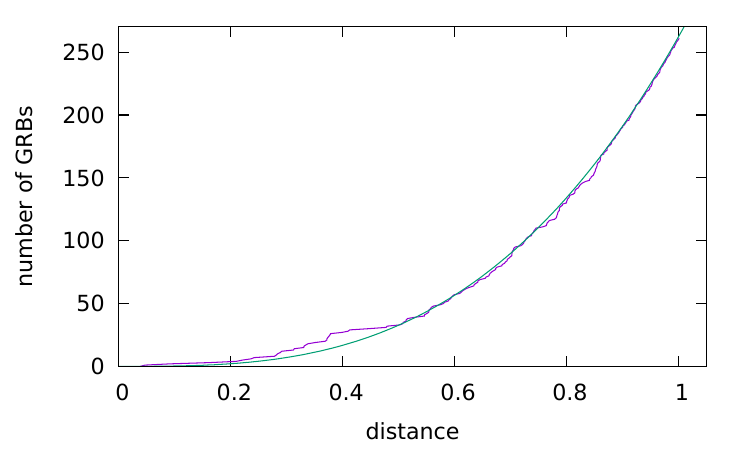}
    \caption{The number of GRBs in the Northern hemisphere, which are closer than a certain transformed distance (ink line). Green line is the $262d^3$ function.}
    \label{fig:koboseszaki}
\end{figure}

A variety of different distance indicators are often used in cosmological measurements (e.g., co-moving distance, luminosity distance). We choose to use the quantity $\log (1+z)$ as a good first-order indicator of the co-moving distance because it is a good approximation of the co-moving distance in the limit of small redshifts. It is valid in the redshift range of our GRB sample ($z \lessapprox 5$).
Using this distance measure, Fig.~\ref{fig:cumdist} demonstrates that the spatial density of GRBs decreases with increasing distance. If the GRB distribution were homogeneous, then the number of objects found within a local volume would increase as the cube of the distance. Instead, a 
 systematic radial density decrease in the number of GRBs per co-moving volume is an indicator of some effect such as a sampling bias and/or source evolution. 
There are two ways in which we might correct for this effect and make the GRBs spatial distribution quasi-homogeneous.
One is to estimate the density around the GRB and use its reciprocal as a weight factor of the GRB. 
We then compute the GRB density within a given volume by summing the weights of the GRBs contained in that volume. 
The other way is to redefine the GRBs' distance so that it creates an approximately third order distance vs. number of object distribution.
We chose to use the latter approach. For the northern galactic hemisphere GRBs distance distribution see Fig.~\ref{fig:koboseszaki}.

This distribution is approximately uniform in distance, and most researchers believe it is isotropic in the sky. One can search for deviations from homogeneity. Searching the less dense part one would always find volumes near the galactic equator that are depleted in GRBs, because the observed afterglows of these GRBs are dimmed by our galaxy. It is much more interesting to search for overdensities. 
Rather than choose random points in the volume, we take a grid with a lattice constant l. We choose another parameter r, which is the radius of a sphere. We put a sphere in every grid points and measure how many GRBs are in this volume of the sphere. We note the maximum of  these numbers K(V). The parameter V is the volume of the sphere goes from 0.001 to 0.5. For bigger V, the sphere typically contains areas where there are no observed GRB, therefore the most dense huge volume finding would contains most of the GRBs.

The probability of measuring the number $K$ for a certain $V$, is obtained by applying the bootstrap method described in our previous works (mnras2024 and \cite{horv25Universe}).
This is done by retaining the sky position of each GRB and replacing its redshift with that of another GRB selected randomly from the sample. 
This is done for all 262 Northern Galactic hemisphere GRBs.  
This process is repeated $W=100$ times so that the number of times $K$ is greater than or equal to the measured value of $K$ can be obtained. 
This frequency estimates the probability of measuring a count of $K$ or larger.
This number is denoted as $P$, and $p=P/W$ estimates the probability. 
Where p was smaller than 0.2 we repeated the process another 100 times (W=200). Where p still was smaller than 0.1 we repeated the process another 100 times (W=300).
Where p still was smaller than 0.06 we repeated the process another 200 times (W=500).
Where p still was smaller than 0.04 we repeated the process another 200 times (W=700).
Where p still was smaller than 0.02 we repeated the process another 300 times, namely 1000 times all together.
Using this so-called progressive scaling we were able to reduce the extremely large computational times which already have cost more than 25 years of single CPU time.

\begin{figure}
    \centering
    \includegraphics[width=0.97\columnwidth]{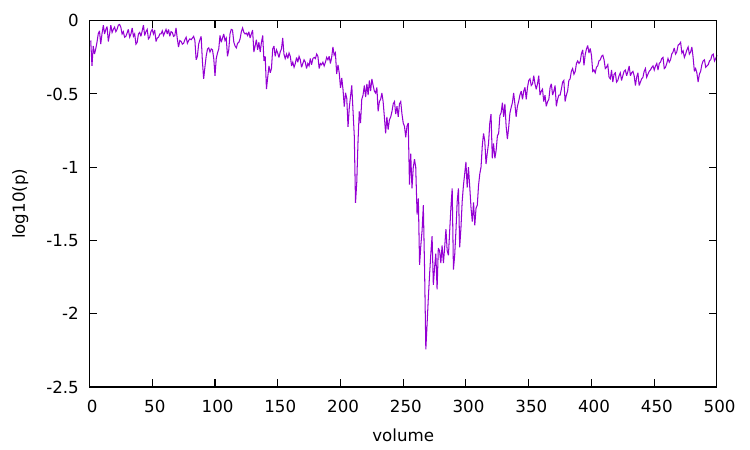}
    \caption{The probabilities to getting higher or equal density by chance vs. volume of the sphere (Northern hemisphere).}
    \label{fig:eszakProb}
\end{figure}

\section{The Northern hemisphere GRB distribution}
\label{sec:eszak}

\begin{figure}
    \centering
    \includegraphics[width=0.97\columnwidth]{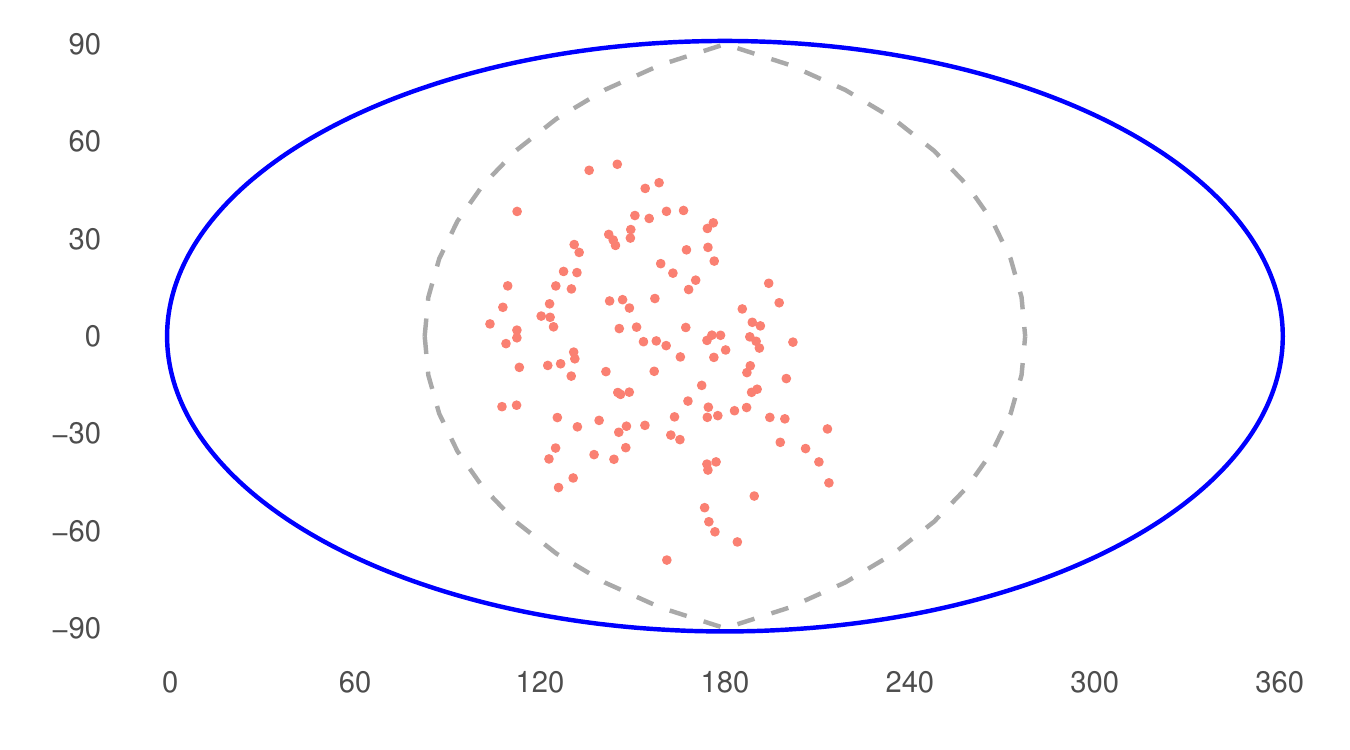}
    \caption{The sky distribution of the 125 GRBs (see the text). The figure is centered on the Northern Galactic Pole. The grey dashed line represents the location of the Galactic equator.}
    \label{fig:eszaksok}
\end{figure}

Fig.~\ref{fig:eszakProb} demonstrates the result of this analysis for the Northern hemisphere. There is a large interval (V=0.255-0.310) where the probability is significant ($p$ less than 0.05).
The most significant part is V=268-270 where the probability is less than 0.01 (with K=125).
Table ~\ref{tab:probok} shows the number of randomization and the number of success getting of K bigger or equal than 125 points (bootstrapped GRBs) for these volumes. 
The probability to getting this deviation by statistical fluctuation is $0.0007 < p < 0.0043$.
Fig.~\ref{fig:eszaksok} shows the sky distribution of the 125 GRBs contained within the most significant overdense sphere. Their sky position consistent with the previously found so called Hercules–Corona Borealis Great Wall.

\begin{table}
    \centering
    \begin{tabular}{l|l|c|c} 
   \multicolumn{1}{c|}{Volume} & 168 & 169 & 170   \\
   \hline
K & 125 & 125  & 125 \\  \hline
W & 1400 & 1400 &  1400  \\  \hline
=125 & 5 & 5 &  8  \\  \hline
>125 & 1 & 4 &  6  \\  \hline
pmin & 0.0007 & 0.0029 &  0.0043  \\  \hline
pmax & 0.0043 & 0.0064 &  0.01 \\   

    \end{tabular}
    \caption{This table shows the 3 highest significant volume in the northern hemisphere. All of them have K=125 (for K see the text). W is the number of the randomized tries. Line 4 contains the number how many times we got 125 among the 1400 tries. Line 5 contains the number how many times we got bigger than 125 among the 1400 tries.  The highest significance is with Volume=168, where the approximate probability is $0.0007 < p < 0.0043$.}
    \label{tab:probok}
\end{table}

\section{The Southern hemisphere GRB distribution}
\label{sec:del}

\label{sec:nhs}

\begin{figure}
    \centering
    \includegraphics[width=0.97\columnwidth]{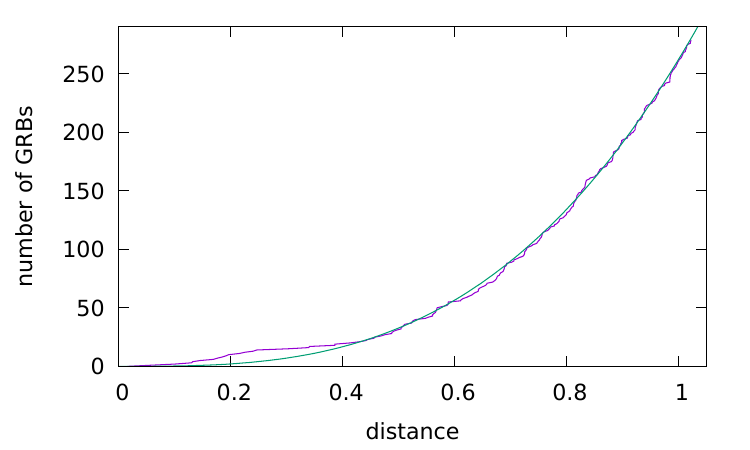}
    \caption{The number of GRBs closer than a certain transformed distance. Southern hemisphere.}
    \label{fig:kobosdeli}
\end{figure}

The southern hemisphere contains 280 GRBs. 
Fig.~\ref{fig:cumdist} and ~\ref{fig:galacticPoleincenter} show the redshift and sky distribution of these 280 GRBs.
For the analysis we used the same method that we used for the northern hemisphere to make the spatial distribution quasi-homogeneous. 
Fig.~\ref{fig:kobosdeli} shows the 280
southern galactic hemisphere GRBs distance distribution.

\begin{figure}
    \centering
    \includegraphics[width=0.97\columnwidth]{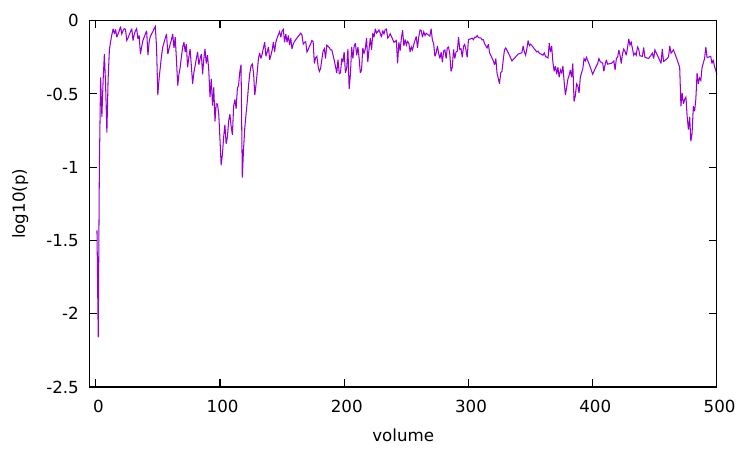}
    \caption{The probabilities (p) to getting higher or equal density by chance vs. volume of the sphere. Southern hemisphere.}
    \label{fig:delProb}
\end{figure}

Similarly to the northern Galactic hemisphere, we search for denser sphere-like volumes in the spatial distribution.
We use the same grid than we used for the northern galactic hemisphere.
Using the same method we got a similar figure than Fig.~\ref{fig:eszakProb}. Fig.~\ref{fig:delProb} shows the probability distribution for the southern galactic hemisphere. 
Only a small area shows significant deviation (p less than 5\%) from homogeneity. 

There are 3 and 4 (this includes the previous 3) GRBs forming a group which is significantly deviates from homogeneity, p=0.037 with 3 and  p=0.007 with 4 GRBs (26 and 7 out of W=700 and 1000). If one includes the closest GRB to this group the probability of getting a 5 GRB cluster is 6\%, therefor the fifth GRB is probably not part of this group. However, it is possible we can get this grouping by statistical fluctuation, with p=0.007. 
Table ~\ref{tab:5grb} shows the redshift and the GRBs distances from each other for these 5 GRBs.
We must point it out that GRB050822 and GRB050318 not just the closest GRBs among the 542 GRBs with known redshift, but they are separated only 5 months in time.
Their angular separation is only 1.03 degree in the sky.
However, an angular separation of $\sim 1^\circ$ is far larger than the image separations expected from gravitational lensing, so a lensing interpretation is very unlikely.

\begin{figure}
    \centering
    \includegraphics[width=7cm]{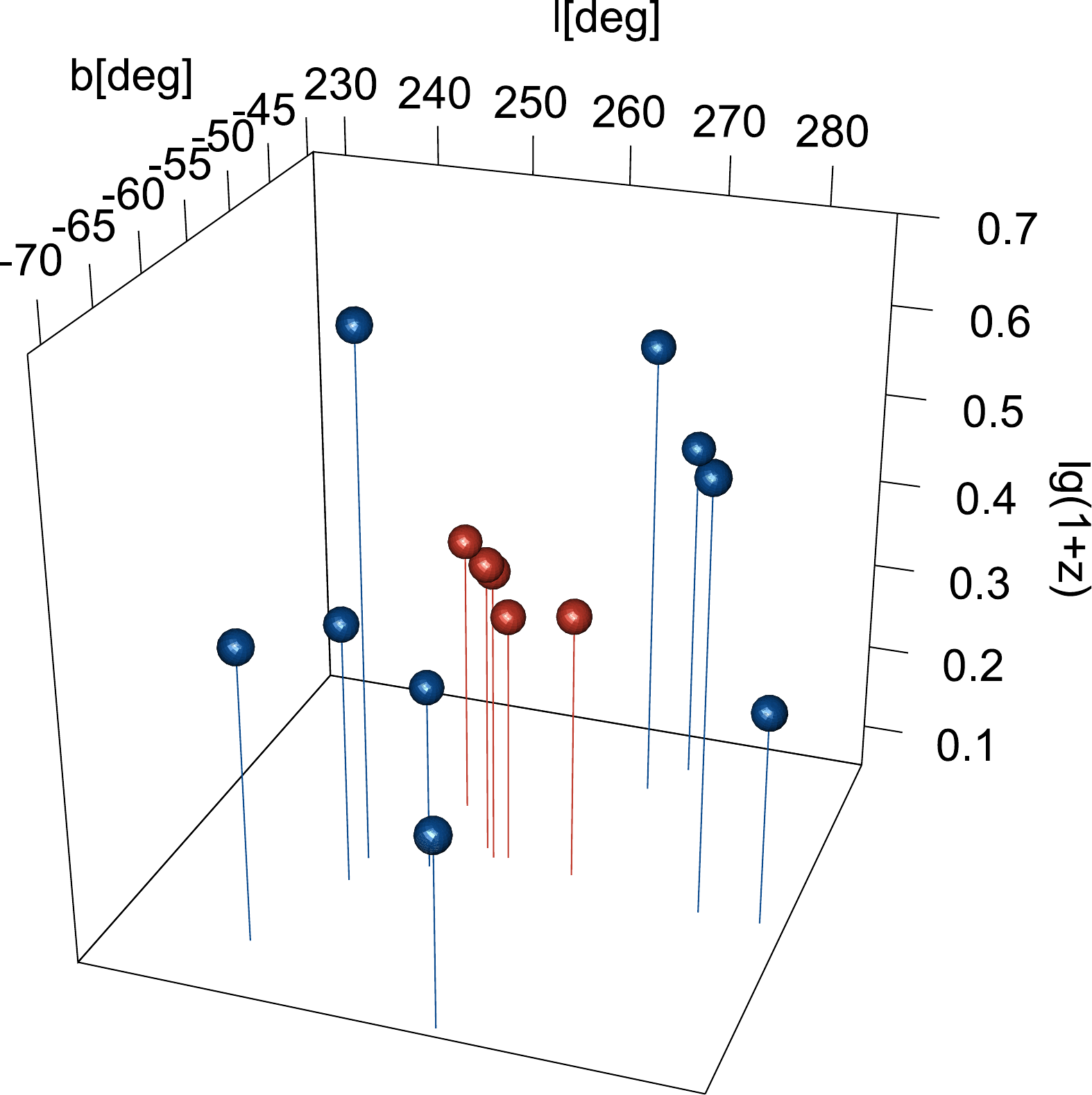}
    \caption{The spatial distribution of the GRB grouping in the southern hemisphere. The identified cluster's members are the inner red marks. A short 3D video can be found in the supplement materials.}
    \label{fig:deli14sp}
\end{figure}

Fig.~\ref{fig:deli14sp} 
shows the spatial distribution of these GRBs.

\begin{table}
    \centering
    \begin{tabular}{l|l|c|c|c|c|c} 
   \multicolumn{1}{c|}{id} & redshift & GRB & GRB & GRB & GRB & GRB \\  
   &&190324&061007&110808&050822&050318\\
   \hline
GRB190324 & 1.1715 & 0 &  &  &  &  \\  \hline
GRB061007 & 1.2622 & 0.0562 & 0 &  &  &  \\  \hline
GRB110808 & 1.3485 & 0.0901 & 0.1355 & 0 &  &  \\  \hline
GRB050822 & 1.434 & 0.0503 & 0.0818 & 0.0679 & 0 &  \\  \hline
GRB050318 & 1.4436 & 0.0482 & 0.0719 & 0.0826 & 0.0148 & 0 \\   

    \end{tabular}
    \caption{This table lists the five GRBs in the Southern hemisphere candidate group, including their redshifts and pairwise separations. }
    \label{tab:5grb}
\end{table}

\section{Discussion}
\label{sec:discussion}

By applying the local volumetric resampling method in a uniform 3D grid framework, we have identified one significant overdensity in the Northern hemisphere and a minor, but potentially interesting, small cluster in the Southern hemisphere.

In the Northern hemisphere, the significant structure was found in a volume corresponding to index values $V = 168$, $169$, and $170$, all producing the same overdensity of $K = 125$ GRBs. The 
estimated probability of such a clustering occurring by chance was $ 0.0007 < p < 0.0043$, strongly rejecting the null hypothesis of homogeneity in this region. The angular distribution of the GRBs in this overdense volume matches previous identifications of the Hercules--Corona Borealis Great Wall, a known large-scale GRB structure. Our results therefore 
confirm this structure's existence and statistical significance using our improved method.

On the Southern hemisphere the overall distribution is more uniform, and only one region was identified (with this paper's method) as having a statistically significant overdensity, with $p$-values close to the 3 $\sigma $ threshold. This cluster consists of a tight group of 4–5 GRBs with small angular separations and closely matched redshifts. Notably, GRB050822 and GRB050318 are separated by only 1.03 degrees and were observed just five months apart, with nearly identical redshifts ($z=1.434$ and $z=1.4436$, respectively). While this temporal and spatial proximity is interesting, the probability of such a grouping occurring by chance is not negligible. 
Furthermore, the angular separation is too large to consider a lensing scenario plausible.

Taken together, these findings suggest anisotropy in the GRB spatial distribution on large scales. The significant structure in the Northern hemisphere —consistent with the HerCrbGW— contrasts sharply with the largely homogeneous distribution in the Southern hemisphere. This asymmetry makes it unlikely that the observed overdensities are merely statistical fluctuations. The absence of multiple significant structures elsewhere in the dataset also strengthens this conclusion. While the local volumetric resampling method is sensitive to overdensities, it did not detect any other statistical anomalies in the GRB spatial distribution.

These results support the idea that GRBs, despite being relatively rare and short-lived events, can serve as tracers of the large-scale structure of the Universe. Moreover, the method used here demonstrates its ability to differentiate between real structures and random fluctuations using both angular and radial information simultaneously. Continued accumulation of GRBs with reliable redshift measurements will further improve the statistical power of such analyses.
It is worth mentioning the high-mass stellar progenitors of gamma-ray bursts, which have measured redshifts. \citet{BalazsRing2015}  have shown that the locations of the universal formation of these objects favor high-mass parts of the cosmic web of dark matter. The large-scale density deviation in the spatial distribution of gamma-ray bursts can be represented by low-frequency spatial harmonics. However, the objects participating in this effect represent only a small fraction of the cosmic matter. Consequently, the density increase found in the spatial distribution of gamma-ray bursts does not necessarily violate the validity of the cosmological principle.

\section{Summary}

We present a statistical analysis of a sample of 542 GRBs with spectroscopic redshifts \citep{2022univ....8..221h, 2022Univ....8..342B}. The local volumetric resampling method was applied to identify statistically significant overdensities in the three-dimensional spatial distribution of these events \citep{hhb14, horv24MNRAS}.

On the Northern Galactic Hemisphere we confirmed   
the so called Hercules–Corona Borealis Great Wall with 125 GRBs.
  
In contrast, the Southern Galactic Hemisphere shows only a marginally significant small group of GRBs, underscoring an intriguing hemispheric asymmetry in the spatial distribution of these events.

The identification of large-scale structures in the Universe has historically relied on observations of luminous tracers. For instance, the Sloan Great Wall, with a characteristic scale of approximately 0.4 Gpc, was identified by galaxy redshift surveys \citep{Gott05}. Even larger coherent structures, such as the Huge Large Quasar Group and the Giant Quasar Arc, each extending over 1.2 Gpc, were subsequently discovered using quasars as cosmic beacons \citep{clo12, 2022MNRASLopez}. The exceptional luminosity of GRBs provides a unique opportunity to probe cosmic architecture on even grander scales.

In linear perturbation of Euclidean cosmological model \cite{Eingorn:2016v} found a $\lambda$ characteristic length of  $\approx\,3700$ Mpc. It is worth
noting  this value  is in the same order as Hercules-Corona Borealis Great Wall, discovered \cite{hhb14}.

This work underscores the necessity of acquiring continued, high-quality spectroscopic redshifts for GRBs. Such data are essential to probe these cosmic megastructures with greater fidelity and to refine our understanding of the Universe's large-scale architecture. Although current observational data are broadly consistent with the large-scale homogeneity of the $\Lambda$CDM model, the detection of highly significant anomalies warrants rigorous investigation with next-generation surveys such as DESI \citep{bottaro.castorina.ea:unveiling}, \textit{Euclid} \citep{euclid-collaboration.scodeggio.ea}, and the \textit{Einstein Probe} mission \citep{Wangetal2024}. These results highlight the unique role of GRBs as luminous, high-redshift tracers, providing an essential complement to galaxy and quasar surveys in mapping the Universe’s largest structures.

\section*{Acknowledgements}
\addcontentsline{toc}{section}{Acknowledgements}
This work was partially supported by the Project Nos. TKP2021-NKTA-64  implemented with funding provided by the Ministry of Culture and Innovation of Hungary from the National Research, Development and Innovation Fund.

\section*{Data Availability}
\label{sec:data}

The data underlying this paper are available in Gamma-Ray Burst Online Index (GRBOX) database published by the Caltech Astronomy Department (\url{https://sites.astro.caltech.edu/grbox/grbox.php}),
Jochen Greiner's table (\url{https://www.mpe.mpg.de/~jcg/grbgen.html}), 
and relevant Gamma-ray Coordination Network (\url{https://gcn.gsfc.nasa.gov/gcn3_archive.html}) messages. The code and data used for this study are available at \url{https://github.com/horvathist/GRBSpherical2025}.



\bibliographystyle{mnras}
\bibliography{horv23}

@preamble{ {\hyphenation{Post-Script Sprin-ger}}
}

@article{clo12,
	title        = {{A structure in the early Universe at z \~ 1.3 that exceeds the homogeneity scale of the R-W concordance cosmology}},
	author       = {{Clowes}, R.~G. and {Harris}, K.~A. and {Raghunathan}, S. and {Campusano}, L.~E. and {S{\"o}chting}, I.~K. and {Graham}, M.~J.},
	year         = 2013,
	month        = {03},
	journal      = {\mnras},
	volume       = 429,
	pages        = {2910--2916},
	doi          = {10.1093/mnras/sts497},
	archiveprefix = {arXiv},
	eprint       = {1211.6256},
	primaryclass = {astro-ph.CO},
	adsurl       = {http://adsabs.harvard.edu/abs/2013MNRAS.429.2910C}
}

@article{Gott05,
	title        = {{A Map of the Universe}},
	author       = {{Gott}, III, J.~R. and {Juri{\'c}}, M. and {Schlegel}, D. and {Hoyle}, F. and {Vogeley}, M. and {Tegmark}, M. and {Bahcall}, N. and {Brinkmann}, J.},
	year         = 2005,
	month        = {05},
	journal      = {\apj},
	volume       = 624,
	pages        = {463--484},
	doi          = {10.1086/428890},
	eprint       = {arXiv:astro-ph/0310571},
	adsurl       = {http://adsabs.harvard.edu/abs/2005ApJ...624..463G}
}

@article{hhb14,
	title        = {{Possible structure in the GRB sky distribution at redshift two}},
	author       = {{Horvath}, I. and {Hakkila}, J. and {Bagoly}, Z.},
	year         = 2014,
	month        = {01},
	journal      = {\aap},
	volume       = 561,
	pages        = {L12},
	doi          = {10.1051/0004-6361/201323020},
	archiveprefix = {arXiv},
	eprint       = {1401.0533},
	primaryclass = {astro-ph.CO},
	eid          = {L12},
	adsurl       = {http://adsabs.harvard.edu/abs/2014A%26A...561L..12H}
}

@article{hbht15,
	title        = {{New data support the existence of the Hercules-Corona Borealis Great Wall}},
	author       = {{Horvath}, I. and {Bagoly}, Z. and {Hakkila}, J. and {T{\'o}th}, L.~V.},
	year         = 2015,
	month        = 12,
	journal      = {\aap},
	volume       = 584,
	pages        = {A48},
	doi          = {10.1051/0004-6361/201424829},
	archiveprefix = {arXiv},
	eprint       = {1510.01933},
	primaryclass = {astro-ph.HE},
	eid          = {A48},
	adsurl       = {http://adsabs.harvard.edu/abs/2015A%26A...584A..48H}
}

@article{HSZ20,
	title        = {{The clustering of gamma-ray bursts in the Hercules-Corona Borealis Great Wall: the largest structure in the Universe?}},
	author       = {{Horvath}, I. and {Sz{\'e}csi}, D. and {Hakkila}, J. and {Szab{\'o}}, {\'A}. and {Racz}, I.~I. and {T{\'o}th}, L.~V. and {Pinter}, S. and {Bagoly}, Z.},
	year         = 2020,
	month        = 10,
	journal      = {\mnras},
	volume       = 498,
	number       = 2,
	pages        = {2544--2553},
	doi          = {10.1093/mnras/staa2460},
	archiveprefix = {arXiv},
	eprint       = {2008.03679},
	primaryclass = {astro-ph.HE},
	adsurl       = {https://ui.adsabs.harvard.edu/abs/2020MNRAS.498.2544H}
}

@article{2022Univ....8..221H,
	title        = {{Does the GRB Duration Depend on Redshift?}},
	author       = {{Horvath}, Istvan and {Racz}, Istvan I. and {Bagoly}, Zsolt and {Bal{\'a}zs}, Lajos G. and {Pinter}, Sandor},
	year         = 2022,
	month        = mar,
	journal      = {Universe},
	volume       = 8,
	number       = 4,
	pages        = 221,
	doi          = {10.3390/universe8040221},
	archiveprefix = {arXiv},
	eprint       = {2206.13621},
	primaryclass = {astro-ph.HE},
	adsurl       = {https://ui.adsabs.harvard.edu/abs/2022Univ....8..221H}
}

@article{horv24MNRAS,
	title        = {{Mapping the Universe with gamma-ray bursts}},
	author       = {{Horvath}, Istvan and {Bagoly}, Zsolt and {Balazs}, Lajos G. and {Hakkila}, Jon and {Horvath}, Zsuzsa and {Joo}, Andras Peter and {Pinter}, Sandor and {T{\'o}th}, L. Viktor and {Veres}, Peter and {Racz}, Istvan I.},
	year         = 2024,
	month        = jan,
	journal      = {\mnras},
	volume       = 527,
	number       = 3,
	pages        = {7191--7202},
	doi          = {10.1093/mnras/stad3669},
	archiveprefix = {arXiv},
	eprint       = {2312.10050},
	primaryclass = {astro-ph.CO},
	adsurl       = {https://ui.adsabs.harvard.edu/abs/2024MNRAS.527.7191H}
}

@article{horv25Universe,
	title        = {{Scanning the Universe for Large-Scale Structures Using Gamma-Ray Bursts}},
	author       = {{Horvath}, Istvan and {Bagoly}, Zsolt and {Balazs}, Lajos G. and {Hakkila}, Jon and {Koncz}, Bendeguz and {Racz}, Istvan I. and {Veres}, Peter and {Pinter}, Sandor},
	year         = 2025,
	month        = apr,
	journal      = {Universe},
	volume       = 11,
	number       = 4,
	pages        = 121,
	doi          = {10.3390/universe11040121},
	eid          = 121,
	archiveprefix = {arXiv},
	eprint       = {2504.05354},
	primaryclass = {astro-ph.CO},
	adsurl       = {https://ui.adsabs.harvard.edu/abs/2025Univ...11..121H}
}

@ARTICLE{2025ApJ989.161Lien,
       author = {{Lien}, Amy Y. and {Krimm}, Hans A. and {Markwardt}, Craig B. and {Oh}, Kyuseok and {Marcotulli}, Lea and {Mushotzky}, Richard and {Collins}, Nicholas R. and {Barthelmy}, Scott D. and {Baumgartner}, Wayne H. and {Cenko}, S. Bradley and {Koss}, Michael and {Laha}, Sibasish and {Sakamoto}, Takanori and {Palmer}, David M. and {Parsotan}, Tyler},
        title = "{The 157 Month Swift/BAT All-sky Hard X-Ray Survey}",
      journal = {\apj},
         year = 2025,
        month = aug,
       volume = {989},
       number = {2},
          eid = {161},
        pages = {161},
          doi = {10.3847/1538-4357/ade676},
archivePrefix = {arXiv},
       eprint = {2506.04109},
 primaryClass = {astro-ph.HE},
       adsurl = {https://ui.adsabs.harvard.edu/abs/2025ApJ...989..161L}
}

@article{li2015,
	title        = {{Testing the homogeneity of the Universe using gamma-ray bursts}},
	author       = {{Li}, Ming-Hua and {Lin}, Hai-Nan},
	year         = 2015,
	month        = 10,
	journal      = {\aap},
	volume       = 582,
	pages        = {A111},
	doi          = {10.1051/0004-6361/201525736},
	eid          = {A111},
	archiveprefix = {arXiv},
	eprint       = {1509.03027},
	primaryclass = {astro-ph.CO},
	adsurl       = {https://ui.adsabs.harvard.edu/abs/2015A&A...582A.111L}
}

@article{2022MNRASLopez,
	title        = {{A Giant Arc on the Sky}},
	author       = {{Lopez}, Alexia M. and {Clowes}, Roger G. and {Williger}, Gerard M.},
	year         = 2022,
	month        = oct,
	journal      = {\mnras},
	volume       = 516,
	number       = 2,
	pages        = {1557--1572},
	doi          = {10.1093/mnras/stac2204},
	archiveprefix = {arXiv},
	eprint       = {2201.06875},
	primaryclass = {astro-ph.CO},
	adsurl       = {https://ui.adsabs.harvard.edu/abs/2022MNRAS.516.1557L}
}

@article{Schulze15,
	title        = {{The Optically Unbiased GRB Host (TOUGH) Survey. VII. The Host Galaxy Luminosity Function: Probing the Relationship between GRBs and Star Formation to Redshift {\ensuremath{\sim}} 6}},
	author       = {{Schulze}, S. and {Chapman}, R. and {Hjorth}, J. and {Levan}, A.~J. and {Jakobsson}, P. and {Bj{\"o}rnsson}, G. and {Perley}, D.~A. and {Kr{\"u}hler}, T. and {Gorosabel}, J. and {Tanvir}, N.~R. and {de Ugarte Postigo}, A. and {Fynbo}, J.~P.~U. and {Milvang-Jensen}, B. and {M{\o}ller}, P. and {Watson}, D.~J.},
	year         = 2015,
	month        = jul,
	journal      = {\apj},
	volume       = 808,
	number       = 1,
	pages        = 73,
	doi          = {10.1088/0004-637X/808/1/73},
	eid          = 73,
	archiveprefix = {arXiv},
	eprint       = {1503.04246},
	primaryclass = {astro-ph.GA},
	adsurl       = {https://ui.adsabs.harvard.edu/abs/2015ApJ...808...73S}
}

@ARTICLE{2012A&A548L7Tello,
       author = {{Tello}, J.~C. and {Castro-Tirado}, A.~J. and {Gorosabel}, J. and {P{\'e}rez-Ram{\'\i}rez}, D. and {Guziy}, S. and {S{\'a}nchez-Ram{\'\i}rez}, R. and {Jel{\'\i}nek}, M. and {Veres}, P. and {Bagoly}, Z.},
        title = "{Searching for Galactic sources in the Swift GRB catalog. Statistical analyses of the angular distributions of FREDs}",
      journal = {\aap},
         year = 2012,
        month = dec,
       volume = {548},
          eid = {L7},
        pages = {L7},
          doi = {10.1051/0004-6361/201220527},
       adsurl = {https://ui.adsabs.harvard.edu/abs/2012A&A...548L...7T}
}

@article{Wang15,
	title        = {{Gamma-ray burst cosmology}},
	author       = {{Wang}, F.~Y. and {Dai}, Z.~G. and {Liang}, E.~W.},
	year         = 2015,
	month        = aug,
	journal      = {\nar},
	volume       = 67,
	pages        = {1--17},
	doi          = {10.1016/j.newar.2015.03.001},
	archiveprefix = {arXiv},
	eprint       = {1504.00735},
	primaryclass = {astro-ph.HE},
	adsurl       = {https://ui.adsabs.harvard.edu/abs/2015NewAR..67....1W}
}

@article{Yad2010,
	title        = {{Fractal dimension as a measure of the scale of homogeneity}},
	author       = {{Yadav}, J.~K. and {Bagla}, J.~S. and {Khandai}, N.},
	year         = 2010,
	month        = {07},
	journal      = {\mnras},
	volume       = 405,
	number       = 3,
	pages        = {2009--2015},
	doi          = {10.1111/j.1365-2966.2010.16612.x},
	archiveprefix = {arXiv},
	eprint       = {1001.0617},
	primaryclass = {astro-ph.CO},
	adsurl       = {http://adsabs.harvard.edu/abs/2010MNRAS.405.2009Y}
}

@article{BalazsTus2018,
	title        = {{Some statistical remarks on the giant GRB ring}},
	author       = {{Bal{\'a}zs}, Lajos G. and {Rejt{\H{o}}}, L{\'\i}dia and {Tusn{\'a}dy}, G{\'a}bor},
	year         = 2018,
	month        = {01},
	journal      = {\mnras},
	volume       = 473,
	number       = 3,
	pages        = {3169--3179},
	doi          = {10.1093/mnras/stx2550},
	archiveprefix = {arXiv},
	eprint       = {1710.01621},
	primaryclass = {astro-ph.CO},
	adsurl       = {https://ui.adsabs.harvard.edu/abs/2018MNRAS.473.3169B}
}

@article{BalazsRing2015,
	title        = {{A giant ring-like structure at 0.78 \&lt; z \&lt; 0.86 displayed by GRBs}},
	author       = {{Bal{\'a}zs}, L.~G. and {Bagoly}, Z. and {Hakkila}, J.~E. and {Horvath}, I. and {K{\'o}bori}, J. and {R{\'a}cz}, I.~I. and {T{\'o}th}, L.~V.},
	year         = 2015,
	month        = {09},
	journal      = {\mnras},
	volume       = 452,
	number       = 3,
	pages        = {2236--2246},
	doi          = {10.1093/mnras/stv1421},
	archiveprefix = {arXiv},
	eprint       = {1507.00675},
	primaryclass = {astro-ph.CO},
	adsurl       = {https://ui.adsabs.harvard.edu/abs/2015MNRAS.452.2236B}
}

@article{Eingorn:2016v,
	title        = {{First-order Cosmological Perturbations Engendered by Point-like Masses}},
	author       = {{Eingorn}, Maxim},
	year         = 2016,
	month        = {07},
	journal      = {\apj},
	volume       = 825,
	number       = 2,
	pages        = 84,
	doi          = {10.3847/0004-637X/825/2/84},
	eid          = 84,
	archiveprefix = {arXiv},
	eprint       = {1509.03835},
	primaryclass = {gr-qc},
	adsurl       = {https://ui.adsabs.harvard.edu/abs/2016ApJ...825...84E}
}

@article{2020ApJ...897..133P,
	title        = {{Cosmicflows-3: The South Pole Wall}},
	author       = {{Pomar{\`e}de}, Daniel and {Tully}, R. Brent and {Graziani}, Romain and {Courtois}, H{\'e}l{\`e}ne M. and {Hoffman}, Yehuda and {Lezmy}, J{\'e}r{\'e}my},
	year         = 2020,
	month        = jul,
	journal      = {\apj},
	volume       = 897,
	number       = 2,
	pages        = 133,
	doi          = {10.3847/1538-4357/ab9952},
	eid          = 133,
	archiveprefix = {arXiv},
	eprint       = {2007.04414},
	primaryclass = {astro-ph.CO},
	adsurl       = {https://ui.adsabs.harvard.edu/abs/2020ApJ...897..133P}
}

@article{2023MNRAS.519L..45S,
	title        = {{King Ghidorah Supercluster: Mapping the light and dark matter in a new supercluster at z = 0.55 using the subaru hyper suprime-cam}},
	author       = {{Shimakawa}, Rhythm and {Okabe}, Nobuhiro and {Shirasaki}, Masato and {Tanaka}, Masayuki},
	year         = 2023,
	month        = feb,
	journal      = {\mnras},
	volume       = 519,
	number       = 1,
	pages        = {L45-L50},
	doi          = {10.1093/mnrasl/slac150},
	archiveprefix = {arXiv},
	eprint       = {2211.11970},
	primaryclass = {astro-ph.GA},
	adsurl       = {https://ui.adsabs.harvard.edu/abs/2023MNRAS.519L..45S}
}

@article{2024JCAP...07..055L,
	title        = {{A Big Ring on the sky}},
	author       = {{Lopez}, A.~M. and {Clowes}, R.~G. and {Williger}, G.~M.},
	year         = 2024,
	month        = jul,
	journal      = {\jcap},
	volume       = 2024,
	number       = 7,
	pages        = {055},
	doi          = {10.1088/1475-7516/2024/07/055},
	eid          = {055},
	archiveprefix = {arXiv},
	eprint       = {2402.07591},
	primaryclass = {astro-ph.CO},
	adsurl       = {https://ui.adsabs.harvard.edu/abs/2024JCAP...07..055L}
}

@article{1991MNRAS.249..218C,
	title        = {{A 100-200 Mpc group of quasars.}},
	author       = {{Clowes}, Roger G. and {Campusano}, Luis E.},
	year         = 1991,
	month        = mar,
	journal      = {\mnras},
	volume       = 249,
	pages        = {218--226},
	doi          = {10.1093/mnras/249.2.218},
	adsurl       = {https://ui.adsabs.harvard.edu/abs/1991MNRAS.249..218C}
}

@article{2012MNRAS.419..556C,
	title        = {{Two close large quasar groups of size {\ensuremath{\sim}} 350 Mpc at z {\ensuremath{\sim}} 1.2}},
	author       = {{Clowes}, Roger G. and {Campusano}, Luis E. and {Graham}, Matthew J. and {S{\"o}chting}, Ilona K.},
	year         = 2012,
	month        = jan,
	journal      = {\mnras},
	volume       = 419,
	number       = 1,
	pages        = {556--565},
	doi          = {10.1111/j.1365-2966.2011.19719.x},
	archiveprefix = {arXiv},
	eprint       = {1108.6221},
	primaryclass = {astro-ph.CO},
	adsurl       = {https://ui.adsabs.harvard.edu/abs/2012MNRAS.419..556C}
}

@article{1997Natur.387..783C,
	title        = {{Discovery of an X-ray afterglow associated with the {\ensuremath{\gamma}}-ray burst of 28 February 1997}},
	author       = {{Costa}, E. and {Frontera}, F. and {Heise}, J. and {Feroci}, M. and {in't Zand}, J. and {Fiore}, F. and {Cinti}, M.~N. and {Dal Fiume}, D. and {Nicastro}, L. and {Orlandini}, M. and {Palazzi}, E. and {Rapisarda\#}, M. and {Zavattini}, G. and {Jager}, R. and {Parmar}, A. and {Owens}, A. and {Molendi}, S. and {Cusumano}, G. and {Maccarone}, M.~C. and {Giarrusso}, S. and {Coletta}, A. and {Antonelli}, L.~A. and {Giommi}, P. and {Muller}, J.~M. and {Piro}, L. and {Butler}, R.~C.},
	year         = 1997,
	month        = jun,
	journal      = {\nat},
	volume       = 387,
	number       = 6635,
	pages        = {783--785},
	doi          = {10.1038/42885},
	archiveprefix = {arXiv},
	eprint       = {astro-ph/9706065},
	primaryclass = {astro-ph},
	adsurl       = {https://ui.adsabs.harvard.edu/abs/1997Natur.387..783C}
}

@inproceedings{racz_akarifis_marton_iau_2016,
	title        = {{A selection of AKARI FIS BSC extragalactic objects}},
	author       = {{Marton}, G. and {T{\'o}th}, L.~V. and {Bal{\'a}zs}, L.~G. and {Zahorecz}, S. and {Bagoly}, Z. and {Horvath}, I. and {R{\'a}cz}, I.~I. and {Nagy}, A.},
	year         = 2016,
	month        = {01},
	booktitle    = {Galaxies at High Redshift and Their Evolution Over Cosmic Time},
	series       = {IAU Symposium},
	volume       = 319,
	pages        = {101--101},
	doi          = {10.1017/S1743921315010297},
	editor       = {{Kaviraj}, S.},
	adsurl       = {https://ui.adsabs.harvard.edu/abs/2016IAUS..319..101M}
}

@article{racz_2018an,
	title        = {Studying the variability of the X-ray spectral parameters of high-redshift GRBs' afterglows},
	author       = {R\'acz, I. I. and Hortobagyi, A. J.},
	year         = 2018,
	journal      = {Astronomische Nachrichten},
	volume       = 339,
	number       = 5,
	pages        = {347--351},
	doi          = {10.1002/asna.201813503},
	url          = {https://onlinelibrary.wiley.com/doi/abs/10.1002/asna.201813503},
	eprint       = {https://onlinelibrary.wiley.com/doi/pdf/10.1002/asna.201813503}
}

@article{2005SSRv..120..143B,
	title        = {{The Burst Alert Telescope (BAT) on the SWIFT Midex Mission}},
	author       = {{Barthelmy}, Scott D. and {Barbier}, Louis M. and {Cummings}, Jay R. and {Fenimore}, Ed E. and {Gehrels}, Neil and {Hullinger}, Derek and {Krimm}, Hans A. and {Markwardt}, Craig B. and {Palmer}, David M. and {Parsons}, Ann and {Sato}, Goro and {Suzuki}, Masaya and {Takahashi}, Tadayuki and {Tashiro}, Makota and {Tueller}, Jack},
	year         = 2005,
	month        = 10,
	journal      = {\ssr},
	volume       = 120,
	number       = {3-4},
	pages        = {143--164},
	doi          = {10.1007/s11214-005-5096-3},
	archiveprefix = {arXiv},
	eprint       = {astro-ph/0507410},
	primaryclass = {astro-ph},
	adsurl       = {https://ui.adsabs.harvard.edu/abs/2005SSRv..120..143B}
}

@article{2004ApJ...611.1005G,
	title        = {{The Swift Gamma-Ray Burst Mission}},
	author       = {{Gehrels}, N. and {Chincarini}, G. and {Giommi}, P. and {Mason}, K.~O. and {Nousek}, J.~A. and {Wells}, A.~A. and {White}, N.~E. and {Barthelmy}, S.~D. and {Burrows}, D.~N. and {Cominsky}, L.~R. and {Hurley}, K.~C. and {Marshall}, F.~E. and {M{\'e}sz{\'a}ros}, P. and {Roming}, P.~W.~A. and {Angelini}, L. and {Barbier}, L.~M. and {Belloni}, T. and {Campana}, S. and {Caraveo}, P.~A. and {Chester}, M.~M. and {Citterio}, O. and {Cline}, T.~L. and {Cropper}, M.~S. and {Cummings}, J.~R. and {Dean}, A.~J. and {Feigelson}, E.~D. and {Fenimore}, E.~E. and {Frail}, D.~A. and {Fruchter}, A.~S. and {Garmire}, G.~P. and {Gendreau}, K. and {Ghisellini}, G. and {Greiner}, J. and {Hill}, J.~E. and {Hunsberger}, S.~D. and {Krimm}, H.~A. and {Kulkarni}, S.~R. and {Kumar}, P. and {Lebrun}, F. and {Lloyd-Ronning}, N.~M. and {Markwardt}, C.~B. and {Mattson}, B.~J. and {Mushotzky}, R.~F. and {Norris}, J.~P. and {Osborne}, J. and {Paczynski}, B. and {Palmer}, D.~M. and {Park}, H. -S. and {Parsons}, A.~M. and {Paul}, J. and {Rees}, M.~J. and {Reynolds}, C.~S. and {Rhoads}, J.~E. and {Sasseen}, T.~P. and {Schaefer}, B.~E. and {Short}, A.~T. and {Smale}, A.~P. and {Smith}, I.~A. and {Stella}, L. and {Tagliaferri}, G. and {Takahashi}, T. and {Tashiro}, M. and {Townsley}, L.~K. and {Tueller}, J. and {Turner}, M.~J.~L. and {Vietri}, M. and {Voges}, W. and {Ward}, M.~J. and {Willingale}, R. and {Zerbi}, F.~M. and {Zhang}, W.~W.},
	year         = 2004,
	month        = {08},
	journal      = {\apj},
	volume       = 611,
	number       = 2,
	pages        = {1005--1020},
	doi          = {10.1086/422091},
	archiveprefix = {arXiv},
	eprint       = {astro-ph/0405233},
	primaryclass = {astro-ph},
	adsurl       = {https://ui.adsabs.harvard.edu/abs/2004ApJ...611.1005G}
}

@article{2011A&A...526A..30G,
	title        = {{The nature of ``dark'' gamma-ray bursts}},
	author       = {{Greiner}, J. and {Kr{\"u}hler}, T. and {Klose}, S. and {Afonso}, P. and {Clemens}, C. and {Filgas}, R. and {Hartmann}, D.~H. and {K{\"u}pc{\"u} Yolda{\textcommabelow s}}, A. and {Nardini}, M. and {Olivares E.}, F. and {Rau}, A. and {Rossi}, A. and {Schady}, P. and {Updike}, A.},
	year         = 2011,
	month        = {02},
	journal      = {\aap},
	volume       = 526,
	pages        = {A30},
	doi          = {10.1051/0004-6361/201015458},
	eid          = {A30},
	archiveprefix = {arXiv},
	eprint       = {1011.0618},
	primaryclass = {astro-ph.HE},
	adsurl       = {https://ui.adsabs.harvard.edu/abs/2011A&A...526A..30G}
}

@article{bottaro.castorina.ea:unveiling,
	title        = {Unveiling Dark Forces with Measurements of the Large Scale Structure of the Universe},
	author       = {Bottaro, Salvatore and Castorina, Emanuele and Costa, Marco and Redigolo, Diego and Salvioni, Ennio},
	year         = 2024,
	month        = {May},
	journal      = {Physical Review Letters},
	volume       = 132,
	pages        = 201002,
	doi          = {10.1103/physrevlett.132.201002},
	url          = {https://link.aps.org/doi/10.1103/PhysRevLett.132.201002}
}

@article{euclid-collaboration.scodeggio.ea,
	title        = {{Euclid preparation. BAO analysis of photometric galaxy clustering in configuration space. I. Covariance matrix estimation and validation of the analysis pipeline}},
	author       = {{Euclid Collaboration} and {Scodeggio}, M. and {Cimatti}, A. and {Amiaux}, J. and {Carbone}, C. and {Castellano}, M. and {Cimatti}, A. and {Conselice}, C.~J. and {Costille}, A. and {D'Alessandro}, M. and {Dore}, O. and {Dubois}, Y. and {Fumana}, M. and {Galeotta}, S. and {H{\"a}ring}, S. and {Jahnke}, K. and {Klaus}, S. and {K{\"o}hlinger}, F. and {Kuntzer}, T. and {Licitra}, R. and {Ligori}, S. and {Maiorano}, L. and {March}, M. and {Martinet}, N. and {Mascia}, F. and {Meinhold}, P.~R. and {Moresco}, M. and {Moscardini}, L. and {Nelson}, A. and {Okumura}, T. and {Parsons}, D. and {Patruno}, A. and {Payne}, A. and {Perez}, E. and {Polenta}, G. and {Potts}, J. and {Rojas}, A. and {Roncarelli}, M. and {Rossetti}, M. and {Sartoris}, B. and {Scarpine}, V. and {Sch{\"a}fer}, B.~M. and {Schneider}, P. and {Sirri}, G. and {Smit}, R. and {Tallada}, M. and {Valiviita}, J. and {Wang}, S. and {Wiesinger}, C. and {Wong}, A. and {Zalesky}, V.},
	year         = 2025,
	month        = mar,
	journal      = {ArXiv},
	pages        = {arXiv:2503.11621},
	eid          = {arXiv:2503.11621},
	archiveprefix = {arXiv},
	eprint       = {2503.11621},
	primaryclass = {astro-ph.CO},
	adsurl       = {https://ui.adsabs.harvard.edu/abs/2025arXiv250311621E}
}

@article{Wangetal2024,
	title        = {Einstein Probe: A new window to the dynamic X-ray Universe},
	author       = {Wang, S. Q. and et al.},
	year         = 2024,
	journal      = {Nature Astronomy}
}

@Article{galaxies13010002,
AUTHOR = {Pe’er, Asaf},
TITLE = {Gamma-Ray Bursts: What Do We Know Today That We Did Not Know 10 Years Ago?},
JOURNAL = {Galaxies},
VOLUME = {13},
YEAR = {2025},
NUMBER = {1},
ARTICLE-NUMBER = {2},
URL = {https://www.mdpi.com/2075-4434/13/1/2},
ISSN = {2075-4434},
DOI = {10.3390/galaxies13010002}
}

@Article{galaxies10030066,
AUTHOR = {Miceli, Davide and Nava, Lara},
TITLE = {Gamma-Ray Bursts Afterglow Physics and the VHE Domain},
JOURNAL = {Galaxies},
VOLUME = {10},
YEAR = {2022},
NUMBER = {3},
ARTICLE-NUMBER = {66},
URL = {https://www.mdpi.com/2075-4434/10/3/66},
ISSN = {2075-4434},
DOI = {10.3390/galaxies10030066}
}

@ARTICLE{2024Univ...10..260F,
       author = {{Frontera}, Filippo},
        title = "{A Short History of the First 50 Years: From the GRB Prompt Emission and Afterglow Discoveries to the Multimessenger Era}",
      journal = {Universe},
         year = 2024,
        month = jun,
       volume = {10},
       number = {6},
          eid = {260},
        pages = {260},
          doi = {10.3390/universe10060260},
archivePrefix = {arXiv},
       eprint = {2407.20305},
 primaryClass = {astro-ph.HE},
       adsurl = {https://ui.adsabs.harvard.edu/abs/2024Univ...10..260F}
}

@ARTICLE{2025RAA....25b5001Y,
       author = {{Yao}, Yu-Hua and {Min}, Fang-Sheng and {Chen}, Shi and {Guo}, Yi-Qing},
        title = "{New Insights on Gamma-Ray Burst Radiation Mechanisms from Multiwavelength Observations}",
      journal = {Research in Astronomy and Astrophysics},
         year = 2025,
        month = feb,
       volume = {25},
       number = {2},
          eid = {025001},
        pages = {025001},
          doi = {10.1088/1674-4527/ad9feb},
archivePrefix = {arXiv},
       eprint = {2412.18577},
 primaryClass = {astro-ph.HE},
       adsurl = {https://ui.adsabs.harvard.edu/abs/2025RAA....25b5001Y}
}

@ARTICLE{2000ApJ...536....1L,
       author = {{Lamb}, Donald Q. and {Reichart}, Daniel E.},
        title = "{Gamma-Ray Bursts as a Probe of the Very High Redshift Universe}",
      journal = {\apj},
         year = 2000,
        month = jun,
       volume = {536},
       number = {1},
        pages = {1-18},
          doi = {10.1086/308918},
archivePrefix = {arXiv},
       eprint = {astro-ph/9909002},
 primaryClass = {astro-ph},
       adsurl = {https://ui.adsabs.harvard.edu/abs/2000ApJ...536....1L}
}

@ARTICLE{2016SSRv..202..195P,
       author = {{Petitjean}, Patrick and {Wang}, F.~Y. and {Wu}, X.~F. and {Wei}, J.~J.},
        title = "{GRBs and Fundamental Physics}",
      journal = {\ssr},
         year = 2016,
        month = dec,
       volume = {202},
       number = {1-4},
        pages = {195-234},
          doi = {10.1007/s11214-016-0235-6},
archivePrefix = {arXiv},
       eprint = {1601.04279},
 primaryClass = {astro-ph.HE},
       adsurl = {https://ui.adsabs.harvard.edu/abs/2016SSRv..202..195P}
}

@INPROCEEDINGS{2012grb..confE..82G,
       author = {{Goldstein}, A.},
        title = "{Difficulties in using GRBs as Standard Candles}",
    booktitle = {Gamma-Ray Bursts 2012 Conference (GRB 2012)},
         year = 2012,
        month = jan,
          eid = {82},
        pages = {82},
          doi = {10.22323/1.152.0082},
       adsurl = {https://ui.adsabs.harvard.edu/abs/2012grb..confE..82G}
}

@incollection{Shawqi_Al_Dallal_2024,
 DOI = {10.9734/bpi/cppsr/v8/11645f},
 author = {Shawqi Al Dallal and Walid J. Azzam},
 pages = {103 - 111},
 publication_type = {incollection},
 title = {Exploring the Origins of Gamma Ray Burst Redshift Distribution in the Early Universe},
 year = {2024},
 month = apr,
 url = {https://hal.science/hal-05238934},
 booktitle = {{Current Perspective to Physical Science Research Vol. 8}},
 publisher = {{BP International}}
}

@ARTICLE{2025MNRAS.542..215L,
       author = {{Liu}, Y. and {Zhang}, Z.~B. and {Dong}, X.~F. and {Li}, L.~B. and {Du}, X.~Y.},
        title = "{The event rate and luminosity function of Fermi/GBM gamma-ray bursts}",
      journal = {\mnras},
         year = 2025,
        month = sep,
       volume = {542},
       number = {1},
        pages = {215-222},
          doi = {10.1093/mnras/staf1217},
archivePrefix = {arXiv},
       eprint = {2507.16595},
 primaryClass = {astro-ph.HE},
       adsurl = {https://ui.adsabs.harvard.edu/abs/2025MNRAS.542..215L}
}

@ARTICLE{2024A&A...685A.163L,
       author = {{Llamas Lanza}, M. and {Godet}, O. and {Arcier}, B. and {Yassine}, M. and {Atteia}, J. -L. and {Bouchet}, L.},
        title = "{High-z gamma-ray burst detection by SVOM/ECLAIRs: Impact of instrumental biases on the bursts' measured properties}",
      journal = {\aap},
         year = 2024,
        month = may,
       volume = {685},
          eid = {A163},
        pages = {A163},
          doi = {10.1051/0004-6361/202347966},
archivePrefix = {arXiv},
       eprint = {2403.03266},
 primaryClass = {astro-ph.HE},
       adsurl = {https://ui.adsabs.harvard.edu/abs/2024A&A...685A.163L}
}

@ARTICLE{2022Univ....8..342B,
       author = {{Bagoly}, Zsolt and {Horvath}, Istv{\'a}n and {Racz}, Istv{\'a}n I. and {Bal{\'a}zs}, Lajos G. and {T{\'o}th}, L. Viktor},
        title = "{The Spatial Distribution of Gamma-Ray Bursts with Measured Redshifts from 24 Years of Observation}",
      journal = {Universe},
         year = 2022,
        month = jun,
       volume = {8},
       number = {7},
          eid = {342},
        pages = {342},
          doi = {10.3390/universe8070342},
       adsurl = {https://ui.adsabs.harvard.edu/abs/2022Univ....8..342B}
}

@ARTICLE{2019MNRAS.486.3027R,
       author = {{{\v{R}}{\'\i}pa}, Jakub and {Shafieloo}, Arman},
        title = "{Update on testing the isotropy of the properties of gamma-ray bursts}",
      journal = {\mnras},
         year = 2019,
        month = jul,
       volume = {486},
       number = {3},
        pages = {3027-3040},
          doi = {10.1093/mnras/stz921},
archivePrefix = {arXiv},
       eprint = {1809.03973},
 primaryClass = {astro-ph.HE},
       adsurl = {https://ui.adsabs.harvard.edu/abs/2019MNRAS.486.3027R}
}

@ARTICLE{2017ApJ...851...15R,
       author = {{{\v{R}}{\'\i}pa}, Jakub and {Shafieloo}, Arman},
        title = "{Testing the Isotropic Universe Using the Gamma-Ray Burst Data of Fermi/GBM}",
      journal = {\apj},
         year = 2017,
        month = dec,
       volume = {851},
       number = {1},
          eid = {15},
        pages = {15},
          doi = {10.3847/1538-4357/aa9708},
archivePrefix = {arXiv},
       eprint = {1706.03556},
 primaryClass = {astro-ph.HE},
       adsurl = {https://ui.adsabs.harvard.edu/abs/2017ApJ...851...15R}
}

@INPROCEEDINGS{2014styd.confE..60B,
       author = {{Bagoly}, Z. and {Balazs}, L.~G. and {Horvath}, I. and {R{\'a}cz}, I. and {Toth}, L.~V. and {Hakkila}, J.},
        title = "{The GRB's Sky Exposure Function}",
    booktitle = {Proceedings of Swift: 10 Years of Discovery (SWIFT 10},
         year = 2014,
        month = dec,
          eid = {60},
        pages = {60},
          doi = {10.22323/1.233.0060},
       adsurl = {https://ui.adsabs.harvard.edu/abs/2014styd.confE..60B}
}

@ARTICLE{2025AcPol..65....9B,
       author = {{Bagoly}, Zsolt},
        title = "{The two-point correlation function of the GRBs}",
      journal = {Acta Polytechnica},
         year = 2025,
        month = mar,
       volume = {65},
       number = {1},
        pages = {9-15},
          doi = {10.14311/AP.2025.65.0009},
       adsurl = {https://ui.adsabs.harvard.edu/abs/2025AcPol..65....9B}
}








\bsp	
\label{lastpage}
\end{document}